 \documentclass[10pt,journal,final,letterpaper,twocolumn,oneside]{IEEEtran}

\usepackage{times,latexsym,amsfonts,amsmath,amssymb,mathrsfs,verbatim,cite}
\usepackage{epsfig,epsf}
\usepackage{graphicx}
\usepackage[dvips]{color}
\usepackage{txfonts}

\def\zZ{{\mathbb Z}}

\def\eE{{\mathbb E}}
\def\pP{{\mathbb P}}

\def\@begintheorem#1#2{\tmpitemindent\itemindent\topsep 0pt\rm\trivlist
    \item[\hskip \labelsep{\indent\it #1\ #2:}]\itemindent\tmpitemindent}
\def\@opargbegintheorem#1#2#3{\tmpitemindent\itemindent\topsep 0pt\rm \trivlist
    \item[\hskip\labelsep{\indent\it #1\ #2\
    \rm(#3):}]\itemindent\tmpitemindent}
\def\@endtheorem{\endtrivlist\unskip}

\newtheorem{theorem}{Theorem}

\newtheorem{remark}{Remark}
\renewcommand{\theequation}{\arabic{section}.\arabic{equation}}

\begin{document}

\title{A Compression Algorithm Using Mis-aligned Side-information$^{\text{\small 1}}$}
\author{\IEEEauthorblockN{Nan Ma, Kannan Ramchandran and David Tse}
\IEEEauthorblockA{\\Wireless Foundations, Dept. of Electrical Engineering and Computer Sciences\\
University of California at Berkeley} }
\maketitle
\begin{abstract}
We study the problem of compressing a source sequence in the presence of side-information that is related to the source via
insertions, deletions and substitutions. We propose a simple algorithm
to compress the source sequence when the side-information is present at both the encoder and decoder.
A key attribute of
the algorithm is that it encodes the edits
contained in runs of different extents separately.
For small insertion and
deletion probabilities, the compression rate of the algorithm is
shown to be asymptotically optimal.
\end{abstract}
\section{Introduction}
\addtocounter{footnote}{+1} \footnotetext{This material is based
upon work supported by the US National Science Foundation (NSF)
under grants 23287 and 30149 and by a gift from Qualcomm Inc.. Any
opinions, findings, and conclusions or recommendations expressed in
this material are those of the authors and do not necessarily
reflect the views of the NSF.}


In \cite{ISIT11}, we have studied the problem of compressing a source
sequence with the help of mis-aligned decoder-only side-information,
where the source and side-information are the input and output of a deletion channel, respectively.
The minimum rate is shown to correspond to the amount of
information in the deleted content plus the locations of the
deletions, minus the uncertainty in the locations given the
source and side-information. We refer to the latter as ``nature's
secret''. This is the information that the encoder and decoder
can never find out. It represents the over-counting of information in the
locations of the deletions. For example, if the input and output of a deletion channel
and are $(0,0)$ and $(0)$, the encoder and decoder will never know and never need to
know whether the first or the second bit is deleted. An interesting question is: how to construct a practical
compression algorithm with the optimal compression rate, where the encoded bits do not reveal ``nature's secret''?
In this paper we provide such a construction for a simpler problem
where the side-information is available at both the encoder and decoder. Although the availability of the side-information is changed,
the minimum rate remains the same.

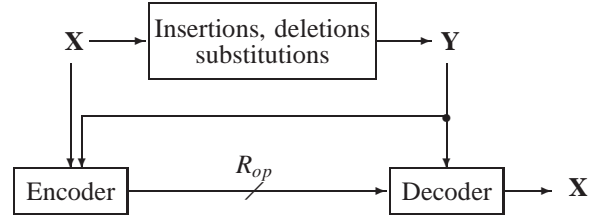
\begin{figure}[!htb]
\centering
\begin{picture}(9,3) 
\put(1,0){
    \put(0,0){\framebox(1.5,0.6){Encoder}}
    \put(5,0){\framebox(1.5,0.6){Decoder}}
    \put(1.8,1.8){\framebox(3,1){$\mbox{Insertions, deletions} \atop \mbox{substitutions}$}}
    \put(0.8,2.3){\makebox(0,0){$\mathbf{X}$}}
    \put(5.8,2.3){\makebox(0,0){$\mathbf{Y}$}}
    \put(7.5,0.35){\makebox(0,0){${\mathbf{X}}$}}
    \put(3.2,0.3){\makebox(0,0){$\diagup$}}
    \put(3.2,0.6){\makebox(0,0){$R_{op}$}}
    \put(0.75,2){\vector(0,-1){1.4}}
    \put(5.75,2){\vector(0,-1){1.4}}
    \put(0.9,1.3){\vector(0,-1){0.7}}
    \put(5.75,1.3){\line(-1,0){4.85}}
    \put(5.75,1.28){\circle*{.1}}
    \put(1.5,0.3){\vector(1,0){3.45}}
    \put(6.5,0.3){\vector(1,0){0.65}}
    \put(1.0,2.3){\vector(1,0){0.75}}
    \put(4.8,2.3){\vector(1,0){0.75}}
   }
\end{picture}
\caption{\label{fig:system} \small \sl Structure of the system}
\end{figure}

In this paper, we study the problem of compressing a source
sequence, $\mathbf{X}$, with the help of side-information, $\mathbf{Y}$, which is available at both the encoder and the decoder. The side-information is related to
the source via insertions, deletions and substitutions.
See Figure~\ref{fig:system} for
an illustration of the system.
The objective of this work is to construct an encoding/decoding algorithm to achieve the optimal compression rate defined as the minimum
number of encoded bits per source bit.

Here is an example of the source and side-information:
\begin{eqnarray*}
\mathbf{X} &=& (0, 0, 1, 1, 0, 1) \\
\mathbf{Y} &=& (0, 1, 0, 0, 1, 1)
\end{eqnarray*}

In order to compare these two sequences, we can insert some gaps, which are denoted by `$-$', to align them as follows.
\begin{eqnarray*}
\mathbf{X}^* &=& (0, 0, 1, 1, 0, 1, -) \\
\mathbf{Y}^* &=& (0, -, 1, 0, 0, 1, 1)
\end{eqnarray*}
This alignment explains the $\mathbf{X}$ with respect to $\mathbf{Y}$ with an insertion,  a substitution and a deletion: $X_2$ is inserted between $Y_1$ and $Y_2$; $X_4$ substitutes $Y_3$; $Y_6$ is deleted. The encoder needs to describe the above editing information using the minimum number of bits.



The problem of synchronizing edited sequences has been studied by
\cite{Leven, OrlitskyDeletion,Trachtenberg06} assuming the number of edits is a
constant that does not increase with the length of the sequence.
Upper and lower bounds on the minimum number of encoded bits
were provided as functions of the number of edits and the length of
the sequence. In \cite{venkataramanan-interactive}, an interactive,
low-complexity and asymptotically optimal scheme was proposed. In
comparison, in this paper,
we consider the case that a fraction of
source bits, rather than a constant number of bits, is edited, which makes the problem more general.
There are also practical synchronization algorithms. such as RSYNC \cite{TridgellRsync} for generic files and VSYNC \cite{vsync}, which targets video applications. In the special case when the source and the side-information differ only by substitutions (side-information is aligned), a universal compression algorithm has been proposed by \cite{Verdu_universal}.

In this paper, we propose a simple compression algorithm, for which the compression rate is asymptotically optimal when the editing probability is small. The key ideas are: (1) describing the locations of insertions and deletions by specifying the runs\footnote{A run is the maximal length sequence of a repeated symbol. The extent, or length, of a run is the number of times the symbol repeats. } of side-information in which they appear, and (2) separately encoding the edits that appears in runs of different extents. To explain idea (1), consider the example where the side-information is $\mathbf{Y}=(0,0,1,0)$ and the source is $\mathbf{X}=(0,1,0)$. Neither the encoder nor the decoder knows whether the first bit or the second bit is deleted. Therefore the encoder needs to describe the location of the deletion only up to a run, which consists of the first two bits in this example, but not further. To explain idea (2), consider the example where the side-information $\mathbf{Y}=(0,0,1,0)$ and the source is $\mathbf{X}=(0,1)$. These sequences can be explained by two deletions, in the first run and the third run of $\mathbf{Y}$, respectively. If the deletion process is memoryless and stationary, the longer first run is more likely to contain a deletion than the shorter third run. Therefore the two deletion events should be encoded separately, using entropy coders with different target distributions, or using a universal entropy coder.

Our compression algorithm can find applications in a number of settings, for example, to compress genomic sequences, as in \cite{Brandon}.\footnote{We would like to thank Dr. Tsachy Weissman
for introducing us to this application.} The difference between the genomic sequences from two individuals of the same species is a small fraction of a whole sequence, and is in the form of insertions, deletions and substitutions. If one of the genomic sequences can be used as side-information, the algorithm can be used to compress the other sequence. The algorithm can also be used in
distributed file backup or file sharing systems, where different source nodes have different versions of the same file differing by a small number of edits including insertions, deletions and substitutions. Here, an old version can be used as side-information that is mis-aligned to the new version of the same file.

The rest of this paper is organized as follows. In
Section~\ref{sec:setup} we formally setup the problem.
In Section~\ref{sec:simplecase} we consider a simple case where the
source sequence is obtained from side-information by pure deletion.
We present the algorithm and analyze the performance.
In Section~\ref{sec:general} we present the algorithm in the general setup.

{\it Notation:} Symbols in boldface represent sequences or matrices, and the symbols in non-boldface represent scalars.
The binary
entropy function is denoted by $h_2(\cdot)$. The notation $\{0,1\}^n$ denotes the $n$-fold Cartesian product
of $\{0,1\}$, and $\{0,1\}^*$ denotes $\left(\bigcup_{k\in \zZ^+}
\{0,1\}^k\right) \bigcup \{\emptyset\}$.

\section{Problem Setup}\label{sec:setup}

We will define two sequences $\mathbf{X}$ and $\mathbf{Y}$, which differ by insertions, deletions, and substitutions.

First, consider an auxiliary length-$n$ sequence $\mathbf{Z}_X=(Z_{X,1},\ldots,Z_{X,n}) \in \{0,1\}^n \sim$ iid Bernoulli$(p)$, where $p\in (0,1)$. Pass $\mathbf{Z}_X$ through a binary symmetric channel with crossover probability $q$ to get $\mathbf{Z}_Y$.

We will then make deletions in $\mathbf{Z}_X$ and $\mathbf{Z}_Y$ to construct $\mathbf{X}$ and $\mathbf{Y}$, respectively.
Let the deletion pattern $\mathbf{D}_X$ be a length-$n$ sequence $\sim$ iid Bernoulli$(d_X)$, which is independent of $\mathbf{Z}_X$ and $\mathbf{Z}_Y$.
The deleted
sequence $\mathbf{X} \in \{0,1\}^*$ is a subsequence of $\mathbf{Z}_X$, which
is derived from $\mathbf{Z}_X$ by deleting all those $Z_{X,i}$'s with $D_{X,i}=1$.
Similarly, the deletion pattern $\mathbf{D}_Y \sim$ iid Bernoulli$(d_Y)$ describes the deletion process from $\mathbf{Z}_Y$ to $\mathbf{Y}$.


Since the editing process from $\mathbf{Z}_X$ to $\mathbf{X}$ is a deletion process, the inverse process from $\mathbf{X}$ to $\mathbf{Z}_X$ can be regarded as an insertion process. Therefore from $\mathbf{X}$ to $\mathbf{Y}$ there are insertions (from $\mathbf{X}$ to $\mathbf{Z}_X$), substitutions (from $\mathbf{Z}_Y$ to $\mathbf{Z}_Y$) and deletions (from $\mathbf{Z}_Y$ to $\mathbf{Y}$).

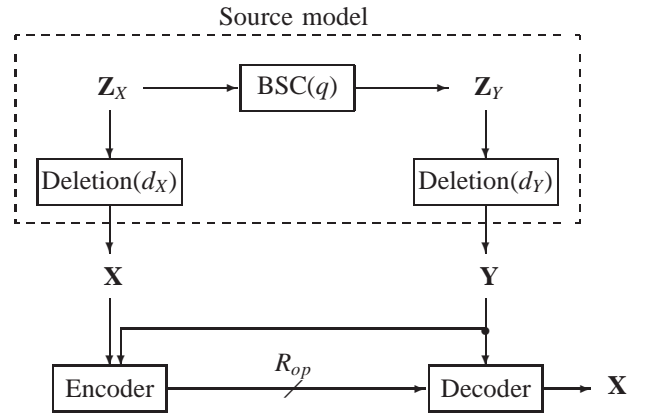
\begin{figure}[!htb]
\centering
\begin{picture}(9,5.2) 
\put(1,0){
    \put(-0.5,2.5){\dashbox{0.1}(7.5,2.5){}}
    \put(0,0){\framebox(1.5,0.6){Encoder}}
    \put(5,0){\framebox(1.5,0.6){Decoder}}
    \put(-0.2,2.75){\framebox(1.9,0.6){Deletion($d_X$)}}
    \put(4.8,2.75){\framebox(1.9,0.6){Deletion($d_Y$)}}
    \put(2.5,4){\framebox(1.5,0.6){BSC($q$)}}
    \put(0.8,4.3){\makebox(0,0){$\mathbf{Z}_X$}}
    \put(5.8,4.3){\makebox(0,0){$\mathbf{Z}_Y$}}
    \put(0.8,1.8){\makebox(0,0){$\mathbf{X}$}}
    \put(5.8,1.8){\makebox(0,0){$\mathbf{Y}$}}
    \put(7.5,0.35){\makebox(0,0){${\mathbf{X}}$}}
    \put(3.2,0.3){\makebox(0,0){$\diagup$}}
    \put(3.2,0.6){\makebox(0,0){$R_{op}$}}
    \put(3.2,5.25){\makebox(0,0){Source model}}
    \put(0.75,4.0){\vector(0,-1){0.65}}
    \put(5.75,4.0){\vector(0,-1){0.65}}

    \put(0.75,2.75){\vector(0,-1){0.65}}
    \put(5.75,2.75){\vector(0,-1){0.65}}

    \put(0.75,1.5){\vector(0,-1){0.9}}
    \put(5.75,1.5){\vector(0,-1){0.9}}
    \put(0.9,1.1){\vector(0,-1){0.5}}
    \put(5.75,1.1){\line(-1,0){4.85}}
    \put(5.75,1.07){\circle*{.1}}
    \put(1.5,0.3){\vector(1,0){3.45}}
    \put(6.5,0.3){\vector(1,0){0.65}}
    \put(1.2,4.3){\vector(1,0){1.25}}
    \put(4.0,4.3){\vector(1,0){1.25}}
   }
\end{picture}
\caption{\label{fig:systemsource} \small \sl Structure of the system with the source model}
\end{figure}

Both sequences $\mathbf{X}$ and $\mathbf{Y}$ are available to the encoder and
$\mathbf{Y}$ is available only to the decoder as
side-information. All the other sequences, $\mathbf{Z}_X$, $\mathbf{Z}_Y$, $\mathbf{D}_X$, and $\mathbf{D}_Y$ are available to neither the encoder nor
the decoder. The encoder encodes $\mathbf{X}$ in the presence of $\mathbf{Y}$ and sends a bit string of variable length to the
decoder so that the decoder can reproduce $\mathbf{X}$ without any error.
The sequences $\mathbf{X}$ and $\mathbf{Y}$ are called the source sequence and the side-information, respectively.
Please see Fig.~\ref{fig:systemsource} for the structure of the system together with the source model.

The performance of the encoder and the decoder is measured by the expected operational rate, which is defined as $R_{op}:=\lim_{n\rightarrow\infty} \eE[L_M/L_Y]$, where $L_M$ is the length of encoded bit string, and $L_Y$ is the length of $Y$.
The objective of this work is to find an encoder and a decoder which minimize the expected operational rate.

\section{Algorithm for the Pure Deletion Case}\label{sec:simplecase}

In order to provide a clear presentation of our algorithm, we start by considering a special case of the general problem, where the source sequence $\mathbf{X}$ is derived from the side-information $\mathbf{Y}$ only by deletion, but not substitution or insertion. Formally speaking, $q=0$ and $d_Y=0$, which imply $\mathbf{Z}_X=\mathbf{Z}_Y=\mathbf{Y}$. For the sake of simplicity, in this section and Appendix~\ref{app:proofthm}, we drop the subscript $X$ in $d_X$ and $\mathbf{D}_X$ and denote them as $d$ and $\mathbf{D}$, respectively.

\subsection{Algorithm for pure
deletion}\label{subsec:algosimplecase}

The encoder has the following three stages.

\begin{enumerate}
  \item Alignment: In this stage we insert some gaps in $\mathbf{X}$ to get $\mathbf{X}^*$, which has the same length as $\mathbf{Y}$. The following greedy alignment algorithm described in \cite[Section~3.1]{MitzSurvey} is used.

      Read $\mathbf{X}$ and $\mathbf{Y}$ from left to right.
Take the first bit of $\mathbf{X}$, and match it with the leftmost appearance of
this bit in $\mathbf{Y}$; then take the second bit of $\mathbf{X}$, and match it with the
subsequent leftmost appearance of this bit in $\mathbf{Y}$; and so on. All the bits in $\mathbf{Y}$ that are not matched with bits from $\mathbf{X}$ are matched with gaps denoted by `$-$'. Let $\mathbf{X}^*$ be the aligned version of $\mathbf{X}$ with gaps inserted. The alignment implies
a reconstructed deletion pattern $\widehat{\mathbf{D}}$, which can explain the deletion process from $\mathbf{Y}$ to $\mathbf{X}$, but is  in general different from $\mathbf{D}$.

  \item Describing the deletions with respect to runs:

  Let
  the maximum extent of the runs in $\mathbf{Y}$ be $L_{\max}$. For IID sequence $\mathbf{Y}$, $\eE[L_{\max}]= \Theta(\log n)$\cite{maxrun}.

The encoder performs the following:

  \begin{itemize}
    \item For $l=1,\ldots, L_{\max}$, do:
    \begin{itemize}
    \item Compute $U_l$, the number of runs of extent $l$ in $\mathbf{Y}$.
    \item For $i= 1,\ldots,U_l$, compute $\widehat{V}_{l,i}$, the number of deletions
     in the $i$-th run of extent
    $l$ in $\mathbf Y$ according to $\widehat{\mathbf{D}}$.
    \end{itemize}
  \end{itemize}

  \item Entropy coding: For each $l=1,\ldots,L_{\max}$, compress the sequence
  $\{\widehat{V}_{l,i}\}_{i=1}^{U_l}$ using an entropy coder. Note that
  $\widehat{V}_{l,i}$ with $l=1,\ldots,L_{\max}$ have different distributions.
\end{enumerate}

The encoded string generated by the encoder is the output of the entropy coder in stage 3).

The decoder has the following two stages.

\begin{enumerate}
  \item Entropy decoder: Reconstruct  $\{\widehat{V}_{i,l}\}_{i=1}^{U_l}$
  for each $l$.
  \item Locate deletions up to runs: For each $l$ and
  each $i$, find the $i$-th run of extent $l$ in $\mathbf Y$, and
  delete $\widehat{V}_{i,l}$ bits in that run. The outcome is the reconstruction of $\mathbf X$.
\end{enumerate}

Since the total number of entries in $\{V_{i,l}\}$ is the total number of runs in $\mathbf{Y}$, which is no larger than $n$, the size of memory the algorithm takes is $O(n)$. Since the greedy alignment, the generation and coding of $\{V_{i,l}\}$ take $O(n)$ operations, the algorithm takes $O(n)$ operations.

\subsection{Example}\label{subsec:examplesimple}
Let the side-information, the hidden deletion pattern, and the source sequence be as follows for example:
\begin{eqnarray*}
\mathbf{Y} &=& (1,0,1,1,0,0,0,1,0,1,1) \\
\mathbf{D} &=& (1,0,0,1,0,1,0,0,0,1,0) \\
\mathbf{X} &=& (0,1,0,0,1,0,1).
\end{eqnarray*}

On the encoder side:

Stage 1): The greedy alignment algorithm aligns $\mathbf{X}$ and $\mathbf{Y}$  and generates $\widehat{\mathbf{D}}$ as follows.
\begin{eqnarray*}
\mathbf{Y} &=& (1,0,1,1,0,0,0,1,0,1,1)\\
\mathbf{X}^* &=& (-,0,1,-,0,0,-,1,0,1,-)\\
\widehat{\mathbf{D}} &=& (1,0,0,1,0,0,1,0,0,0,1).
\end{eqnarray*}

Stage 2): The maximum extent of the runs in $\mathbf{Y}$ is $L_{\max}=3$. There are $U_1=4$ runs of extent $1$, $U_2=2$ runs of extent $2$, and $U_3=1$ run of extent $3$. For the four extent-$1$ runs, `$1$', `$0$', `$1$' and `$0$', only the first one is deleted according to $\widehat{\mathbf{D}}$, therefore
we have
\[(V_{1,1},V_{1,2},V_{1,3},V_{1,4}) = (1,0,0,0).\]

For the two extent-$2$ runs, `$1,1$' and `$1,1$', there is a deletion in each of them. Therefore we have
\[(V_{2,1},V_{2,2}) = (1,1).\]

For the only extent-$3$ run, `$0,0,0$', there is a deletion in it. Therefore we have
\[(V_{3,1}) = (1).\]

Stage 3):
The entropy encoder compresses $((V_{1,1},V_{1,2},V_{1,3},V_{1,4}), (V_{2,1},V_{2,2}), (V_{3,1}))=((1,0,0,0),(1,1),(1))$. Note that each entry in $(V_{1,1},V_{1,2},V_{1,3},V_{1,4})$ is more likely to be $0$ than $(V_{2,1},V_{2,2})$ and $(V_{3,1})$. Therefore we should use entropy encoder with different target distributions to encode them, when the sequences are long.

On the decoder side:

Stage 1): The entropy decoder reconstructs $((V_{1,1},V_{1,2},V_{1,3},V_{1,4}), (V_{2,1},V_{2,2}), (V_{3,1}))=((1,0,0,0),(1,1),(1))$.

Stage 2): Since $(V_{1,1},V_{1,2},V_{1,3},V_{1,4}) = (1,0,0,0)$, the decoder deletes the first run of extent-$1$, i.e., the first bit. Since
$(V_{2,1},V_{2,2}) = (1,1)$, the decoder deletes a bit from each of the two runs of extent-$2$. It does not matter which bit to delete in each run. Since $(V_{3,1}) = (1)$, the decoder deletes a bit in the only extent-$3$ run. The deletions are represented by $\widetilde{\mathbf{D}}$ and the reconstruction of the source sequence is denoted by $\widetilde{\mathbf{X}}$.
\begin{eqnarray*}
\mathbf{Y} &=& (1,0,1,1,0,0,0,1,0,1,1)\\
\widetilde{\mathbf{D}} &=& (1,0,1,0,0,0,1,0,0,0,1)\\
\widetilde{\mathbf{X}} &=& (0,1,0,0,1,0,1).
\end{eqnarray*}

Since $\widetilde{\mathbf{X}} = {\mathbf{X}}$, the reconstruction is correct.

\subsection{Performance of the algorithm}
Let $\mathbf U:=\{U_l\}_{l=1}^{L_{\max}}$ and $\widehat{\mathbf
{V}}:=\{\widehat{V}_{l,i}\}_{l=1,i=1}^{L_{\max},U_l}$. In the limit as the lengths of the sequences tends to infinity, the operational rate of this algorithm is
$R_{op}=\lim_{n\rightarrow \infty}H(\widehat{\mathbf {V}})/n$. The optimal rate is $\lim_{n\rightarrow \infty}H(\mathbf{X}|\mathbf{Y})/n$.
When the probability of deletion $d$ is small, the following theorem shows that the algorithm is asymptotically optimal.

\begin{theorem}\label{thm:gapiid}
The gap between the operational rate of the algorithm described in Section~\ref{subsec:algosimplecase} and the optimal rate satisfies:
$\lim_{n\rightarrow \infty} [H(\widehat{\mathbf {V}})/n - H(\mathbf{X} | \mathbf{Y})/n] = O(d^{2-\epsilon})$, for any $\epsilon >0$.
\end{theorem}

The proof is provided in Appendix~\ref{app:proofthm}, which can be intuitively explained as follows. When $d$ is small, the deletions are typically far away from each other. Therefore the intervals between the deletions are so long that can be used to synchronize segments of $\mathbf{X}$ to segments of $\mathbf{Y}$. As a result, the deletions can be located within the correct runs with high probability.
The exact positions of the deletions within the runs are impossible to find based on only $\mathbf{X}$ and $\mathbf{Y}$. Since the goal is to reconstruct $\mathbf{X}$, describing the positions within runs is unnecessary. Moreover, the description of the locations of the deletions, $\widehat{\mathbf{V}}$, is almost independent of the decoder side-information $\mathbf{Y}$. Therefore sending $\widehat{\mathbf{V}}$ is approximately optimal in terms of rate. See Section~\ref{subsec:comparison}-2 for more discussions about the independence between $\widehat{\mathbf{V}}$ and $\mathbf{Y}$. The deletions cannot be located within the correct runs only if two or more deletions are in the same run or adjacent runs, which occurs with the probability in the order of $O(d^2)$. Therefore the gap between the operational rate and the optimum is in the order of $O(d^{2-\epsilon})$.

\begin{remark}
In \cite{ISIT11}, we have shown that when $p=1/2$, for any $\epsilon >0$, $\lim_{n\rightarrow \infty} H(\mathbf{X}|\mathbf{Y})/n =  h_2(d) - c  d + O(d^{2-\epsilon})$, where $c:=\sum_{l=1}^{\infty}2^{-l-1}l \log_2 l \approx 1.29$.\footnote{In \cite{ISIT11}, $\mathbf{Y}$ is defined as the deleted version of $\mathbf{X}$. Therefore the expression $H(\mathbf{X}|\mathbf{Y})$ in this paper corresponds to $H(\mathbf{Y}|\mathbf{X})$ in \cite{ISIT11}.} It captures the asymptotic expansion of the optimal rate to the precision of $\Theta(d)$ with a remainder term $O(d^{2-\epsilon})$. Due to Theorem~\ref{thm:gapiid}, $R_{op} =  h_2(d) - c  d + O(d^{2-\epsilon})$, which also matches the optimal rate to the precision of $\Theta(d)$.
\end{remark}

\begin{remark}
In \cite{ISIT11}, we have shown that $\lim_{n\rightarrow \infty}H(\mathbf{X}|\mathbf{Y})/n$ is also the minimum rate when the side-information is only available available at the decoder but not the encoder. Although the minimum rate is the same, constructing an explicit algorithm to implement the distributed compression at the asymptotically optimal rate remains an open problem.
\end{remark}

\subsection{Comparison to other compression algorithms}\label{subsec:comparison}

Let us compare the algorithm described in Section~\ref{subsec:algosimplecase} with two simpler but suboptimal algorithms in the simple case $\mathbf{Y}\sim$ iid Bernoulli$(1/2)$ ($p=1/2$). The comparison reveals more intuition on why the algorithm is asymptotically optimal.

\subsubsection{Sending $\widehat{\mathbf{D}}$ directly}
A simple and the most natural algorithm to compress $\mathbf{X}$ given $\mathbf{Y}$ is first running a greedy alignment to obtain $\widehat{\mathbf{D}}$ (as in stage 1)) and then compressing $\widehat{\mathbf{D}}$ using an entropy coder (similar to stage 3)). As the lengths of the sequences tend to infinity, the operational rate is $\lim_{n\rightarrow \infty} H(\widehat{\mathbf{D}})/n$. If we approximate $H(\widehat{\mathbf{D}})$ by $H({\mathbf{D}})$\footnote{It can be made rigorous using the techniques that are similar to those used to prove argument (iii) in Appendix~\ref{app:proofthm}}, the operational rate is approximately $h_2(d) = -d \log_2 d + d \log_2 e + O(d^2)$. Therefore for small $d$, the operational rate of this simple algorithm matches the optimal expression up to the $-d \log_2 d$ term.
But for the $\Theta(d)$ term, there is a gap $c d \approx 1.29 d$. That is, this compression algorithm wastes $1.29$ bits per deletion bit on average. When $d$ is not very small, $-d \log_2 d$ and $d$ can be in the same order of magnitude. Therefore the gap may not be negligible in practice.

The above strategy is suboptimal because $\widehat{\mathbf{D}}$ specifies the exact positions of the deletions. Note that after specifying the runs that contain the deletions and specifying the number of deletions in each run, $\mathbf{X}$ can already be deduced from $\mathbf{Y}$. However, this strategy goes further and specifies the exact positions within the runs, which are redundant in terms of reconstructing $\mathbf{X}$. Therefore this strategy over-describes the positions of the deletions beyond what is necessary to represent $\mathbf{X}$. The amount of over-description, $H(\mathbf{D}|\mathbf{X}, \mathbf{Y})$, is called ``nature's secret'' in \cite{ISIT11}, because only the hypothetical party ``nature'' has access to $\mathbf{D}$, but the encoder and decoder do not.

\subsubsection{Locating deletions up to runs}\label{subsec:comparison2}
The analysis of the previous strategy suggests that the encoder should specify the location of the deletions with respect to runs. Therefore a better algorithm than the one described in Section~\ref{subsec:comparison}-1 is first defining a sequence $\widehat{\mathbf{W}}$ such that $\widehat{W}_i$ is the number of deletions in the $i$-th run of $\mathbf{Y}$ according to $\widehat{\mathbf{D}}$, then compressing $\widehat{\mathbf{W}}$ at the entropy rate.

Since the average extent of a run in an iid Bernoulli$(1/2)$ sequence is $2$, the length of $\widehat{\mathbf{W}}$ is approximately half of that of $\widehat{\mathbf{D}}$. It can be shown\footnote{Using the techniques that are similar to those used to prove argument (iii) in Appendix~\ref{app:proofthm}} that the operational rate can be approximated by $(h_2(d) - d)$. There is still a linear $d$ gap between this rate and the optimal one, given by $(c-1)d \approx 0.29 d$. That is, this algorithm wastes $0.29$ bit per deletion bit.

Why is this algorithm suboptimal? The reason is because $\widehat{\mathbf{W}}$ is significantly correlated with $\mathbf{Y}$. If the deletion process is iid, then the longer runs of $\mathbf{Y}$ tend to contain more deletions and the shorter runs tend to contain less deletions. Therefore $\mathbf{Y}$ reveals a certain amount of information about $\widehat{\mathbf{W}}$, that is about $0.29$ bit per deletion bit. The algorithm described above does not use this amount of information and thus is suboptimal.

The algorithm described in Section~\ref{subsec:algosimplecase}, however, treats the deletions contained in runs of different extents differently. As a result the operational rate matches the optimal rate for the $\Theta(d)$ term.

Table~\ref{tab:comparison} provides a comparison among the performance of the two algorithms in Section~\ref{subsec:comparison} and the one in Section~\ref{subsec:algosimplecase} for $n=1000$kb and $d=0.01$. Note that when $\mathbf{Y}$ has biased bits ($p=0.1$), the benefit of the proposed algorithm in Section~\ref{subsec:algosimplecase} is more significant than when $p=0.5$. The reason is that when $p=0.1$, the runs of $\mathbf{Y}$ are longer and it pays to exploit the information from the run-lengths.

\begin{table}
\caption{\label{tab:comparison} \small \sl Performance of compression algorithms for $n=1000$kb, $d=0.01$.}
\centering
\begin{tabular}{|c|c|c|c|c|}
  \hline
   p & No SI & Sec. \ref{subsec:comparison}-1 & Sec. \ref{subsec:comparison}-2 & Sec. \ref{subsec:algosimplecase} \\
   \hline
  0.5 & 990kb & 81kb & 71kb & 68kb \\
  0.1 & 469kb & 81kb & 63kb & 46kb \\
  \hline
\end{tabular}
\end{table}

\section{Algorithm for the General Case}\label{sec:general}

The algorithm described in Section~\ref{subsec:algosimplecase} can be extended to the general problem where $\mathbf{Y}$ is related to $\mathbf{X}$ by insertions, deletions and substitutions.

\subsection{Algorithm for insertions, deletions and substitutions}\label{subsec:algogeneralcase}

The encoder has the following stages.

\begin{enumerate}
  \item Alignment: align $\mathbf{X}$ and $\mathbf{Y}$ using the minimum total number of insertions, deletions and substitutions. If there are multiple such alignments, pick any one of them. This can be done by the Needleman-Wunsch algorithm \cite{nwalign} with the gap penalty and the substitution penalty equal to $1$, with computation complexity of order $O(n^2)$.
      The algorithm generates two sequences $\mathbf{X}^*$ and $\mathbf{Y}^*$, which are $\mathbf{X}$ and $\mathbf{Y}$ with gaps, respectively. Then construct $\widehat{\mathbf{Z}}_X$ and $\widehat{\mathbf{Z}}_Y$ by replacing the gaps in $\mathbf{X}^*$ and $\mathbf{Y}^*$ by the corresponding bits in $\mathbf{Y}^*$ and $\mathbf{X}^*$, respectively.

  \item Describing the insertions (from $\mathbf{Y}$  to $\widehat{\mathbf{Z}}_Y$):

The edits from $\mathbf{Y}$  to $\widehat{\mathbf{Z}}_Y$ can be viewed as insertions. The locations of the insertions are specified by the gaps in $\mathbf{Y}^*$. The content of the insertions is specified by the corresponding bits in $\widehat{\mathbf{Z}}_Y$.

All the insertions can be categorized into isolated insertions with only one bit per insertion event, and bursts of insertions with two or more consecutive bits per insertion event. For each insolated insertion, if the inserted bit is equal to the bit on the left (or right) side, the insertion is extending the run to the left (or right). If the inserted bit is not equal to the bits on either side, it is breaking an existing run and creating a new run. We will describe the isolated insertions that extend runs, then the insertions that break runs, then the bursts of insertions.

\begin{itemize}
  \item In order to describe the insertions that extend runs, the encoder does the following.
\begin{itemize}
    \item For $l=1,\ldots, L_{\max}$ ($L_{\max}$ is the the maximum extent of the runs in $\mathbf{Y}$), do:
    \begin{itemize}
    \item For $i= 1,\ldots,U_l$ ($U_l$ is the number of runs of extent $l$ in $\mathbf{Y}$), let $\widehat{V}_{l,i}^{ins}:= 1$ if
    the $i$-th run of extent $l$ in $\mathbf Y$ is extended by one bit, and $\widehat{V}_{l,i}^{ins}:= 0$ otherwise.
    \end{itemize}
\end{itemize}
Having made such insertions, $\mathbf{Y}$ becomes $\mathbf{Y}'$.
  \item In order to describe the insertions that break runs, the encoder does the following.

  In the sequence $\mathbf{Y}'$, a slot between two bits is a potential location to break a run only if the two bits are the same. The slots before the first bit and after the last bit are also potential locations to create new runs. Let $U_0$ denote the total number of such potential locations in $\mathbf{Y}'$. For $i=1,\ldots,U_0$, let $\widehat{V}_{0,i}^{ins}:= 1$ if a bit is inserted in the $i$-th potential location, and $\widehat{V}_{0,i}^{ins}:= 0$ otherwise.

Having made such insertions, $\mathbf{Y}'$ becomes $\mathbf{Y}''$. Let $\widehat{\mathbf{V}}^{ins}$ denote all the descriptions up to this step: $\{\widehat{V}_{l,i}^{ins}\}_{l\geq 0}$.

  \item In order to describe the bursts of insertions, the encoder creates a sequence $\widehat{\mathbf{V}}^{burst}$ from $\widehat{\mathbf{Z}}_Y$ by keeping the bursts of inserted bits and replacing the other bits by `$*$'. $\widehat{\mathbf{V}}^{burst}$ describes the insertions needed to construct $\widehat{\mathbf{Z}}_Y$ from $\mathbf{Y}''$.
\end{itemize}

\item Describe the substitutions (from $\widehat{\mathbf{Z}}_Y$  to $\widehat{\mathbf{Z}}_X$):

The edits from $\widehat{\mathbf{Z}}_Y$  to $\widehat{\mathbf{Z}}_X$ can be viewed as substitutions, which can be described by $\widehat{\mathbf{V}}^{sub}:=\widehat{\mathbf{Z}}_Y\oplus \widehat{\mathbf{Z}}_X$.

\item Describe the deletions (from $\widehat{\mathbf{Z}}_X$  to $\mathbf{X}$) as in stage 2) of Section~\ref{subsec:algosimplecase}. Denote the description by $\widehat{\mathbf{V}}^{del}$.

\item Entropy coding: Use an entropy coder to compress
  $\widehat{\mathbf{V}}^{ins}$, $\widehat{\mathbf{V}}^{burst}$, $\widehat{\mathbf{V}}^{sub}$ and $\widehat{\mathbf{V}}^{del}$.
\end{enumerate}

The decoder decodes  $\widehat{\mathbf{V}}^{ins}$, $\widehat{\mathbf{V}}^{burst}$, $\widehat{\mathbf{V}}^{sub}$ and $\widehat{\mathbf{V}}^{del}$ by an entropy decoder, and then follow the stages 2) to 4) to construct $\mathbf{X}$ from $\mathbf{Y}$.

\subsection{Performance analysis}

The operational rate of the above algorithm can be analyzed for small probability of insertion, deletion and substitution as follows.

\begin{theorem}\label{thm:gapgeneral}
The gap between the operational rate of the algorithm described in Section~\ref{subsec:algogeneralcase} and the optimal rate satisfies:
$\lim_{n\rightarrow \infty} [H(\widehat{\mathbf {V}}^{ins},\widehat{\mathbf {V}}^{burst}, \widehat{\mathbf {V}}^{sub}, \widehat{\mathbf {V}}^{del})/n - H(\mathbf{X} | \mathbf{Y})/n] = O(d^{2-\epsilon})$, for any $\epsilon >0$, where $d=\max\{ d_X, d_Y, q\}$.
\end{theorem}

The proof is similar to that of Theorem~\ref{thm:gapiid} and is provided in Appendix~\ref{app:proofthmgeneral}.

Intuitively, when the editing probabilities $d_X$, $d_Y$ and $q$ are small, the edits are typically far away from each other. Therefore the intervals between the edits are so long that the segments of $\mathbf{X}$ in the intervals can be correctly matched to the corresponding segments of $\mathbf{Y}$. As a result, the edits can be isolated. The operational message rate is approximately equal to the summation of the message rates in the pure deletion problem, the pure insertion problem and pure substitution problem.
On the other hand, the conditional entropy rate $\lim_{n\rightarrow \infty} H(\mathbf{X}|\mathbf{Y})$ can be also approximated by the conditional entropy rates of the pure deletion problem ($\lim_{n\rightarrow \infty} H(\mathbf{X}|\mathbf{Z}_X)$), the pure substitution problem ($\lim_{n\rightarrow \infty} H(\mathbf{Z}_X|\mathbf{Z}_Y)$), and the pure insertion problem ($\lim_{n\rightarrow \infty} H(\mathbf{Z}_Y|\mathbf{Y})$), with an approximation gap no more than $O(d^{2-\epsilon})$. Therefore the algorithm described above is asymptotically optimal.

\section{Concluding Remarks}\label{sec:conclusion}

We have studied
the problem of compressing a source sequence in the presence of side-information that is mis-aligned to the source due to
insertions, deletions and substitutions. We have proposed an algorithm
to compress the source sequence given the side-information at both the encoder and decoder.
For small insertion and
deletion probability, the compression rate of the algorithm is
asymptotically optimal.
Directions for future work include (1) developing algorithms for bursty insertions, deletions, and substitutions, and (2) developing distributed algorithms to compress a source sequence when the reference sequence is only available at the decoder side.

\appendices
\renewcommand{\theequation}{\thesection.\arabic{equation}}
\setcounter{equation}{0}


\section{Proof of Theorem~\ref{thm:gapiid}}\label{app:proofthm}

Stage 2) of the algorithm described in Section~\ref{subsec:algosimplecase} compresses the reconstructed deletion pattern $\widehat{\mathbf{D}}$  and  to generate $\widehat{\mathbf{V}}$. Let $\mathbf{V}$ be the output if the true deletion pattern $\mathbf{D}$ would be used as the input. Note that the sizes of $\mathbf{V}$ and $\widehat{\mathbf{V}}$ are identical, because they are both determined by $\mathbf{Y}$.

We have
\begin{eqnarray*}
H(\mathbf X|\mathbf Y) &=& H(\mathbf X, \mathbf V|\mathbf Y) -
H(\mathbf V|\mathbf X, \mathbf Y)\\
&\stackrel{(a)}{=}& H(\mathbf V|\mathbf Y) -
H(\mathbf V|\mathbf X, \mathbf Y)\\
&=& H(\mathbf V) - I(\mathbf V;\mathbf Y)- H(\mathbf V|\mathbf X,
\mathbf Y),\\
&=& H(\widehat{\mathbf{V}}) - I(\mathbf V;\mathbf Y)- H(\mathbf V|\mathbf X,
\mathbf Y)-(H(\widehat{\mathbf{V}})-H(\mathbf{V})),
\end{eqnarray*}
where step (a) is because
$\mathbf X$ is determined by $\mathbf
Y$ and $\mathbf V$.
We will prove the following three arguments: (i)$\lim_{n\rightarrow \infty} I(\mathbf V;\mathbf Y)/n=0$,
(ii) $\lim_{n\rightarrow \infty} I(\mathbf V|\mathbf X, \mathbf Y)/n = O(d^{2-\epsilon})$ for any $\epsilon >0$, and (iii)
$\lim_{n\rightarrow \infty} |H(\widehat{\mathbf{V}})-H(\mathbf{V})|/n = O(d^{2-\epsilon})$ for any $\epsilon >0$.
\\

\noindent\emph{Proof of argument (i):} Given $U_l=u_l$, $\{{V}_{l,i}\}_{i=1}^{U_l}$ is an iid sequence with distribution $p_{V_{l,i}}(v) = {l \choose v} d^v (1-d)^{l-v}$. Therefore $\mathbf{Y} - \mathbf U - \mathbf
V$ forms a Markov chain. By the data processing inequality, we have $\lim_{n\rightarrow \infty}
I(\mathbf{X};\mathbf{V})/n \leq \lim_{n\rightarrow \infty} H(\mathbf{U})/n =
0$. Therefore (i) is proved.
\\

\noindent\emph{Proof of argument (ii):}
Let an extended run be a run along with one additional bit at each end of
the run \cite{MontanariISIT10}. We call an extended run of $\mathbf{Y}$ atypical if it contains more than one deletion according to $\mathbf{D}$. Let $\mathbf{D}^*$ be the sequence that is identical to $\mathbf{D}$ in the atypical extended runs, and is equal to `$*$' otherwise. Suppose there are $K$ runs of  `$*$'s in $\mathbf{D}^*$, and the $i$-th run starts from position $a_i$ and ends at position $b_i$. Let $C_i:= \sum_{j = a_i}^{b_i} D_j$. Let $\mathbf{C}:= (C_1,\ldots,C_K)$.

With the help of $\mathbf{D}^*$ and $\mathbf{C}$, aligning $\mathbf{X}$ and $\mathbf{Y}$ becomes easier. The atypical extended runs divide the whole sequences into $K$ segments. One can locate $K$ segments in $\mathbf{X}$, each of which corresponds to a run of `$*$'s in $\mathbf{D}^*$. Within each segment, there are no longer atypical extended runs, and the deletions can be located in the correct runs without any ambiguity \cite[Proof of Lemma~IV.4]{MontanariISIT10}. Since $\mathbf{V}$ is only about the locations of deletions up to runs, $H(\mathbf{V}|\mathbf{X}, \mathbf{Y}, \mathbf{D}^*, \mathbf{C})=0$, which implies that $H(\mathbf{V}|\mathbf{X}, \mathbf{Y})\leq H(\mathbf{D}^*, \mathbf{C})$. Since an extended run is atypical with probability $O(d^2)$, $\eE[K]/n = O(d^2)$ and $H(\mathbf{D}^*, \mathbf{C})/n = O(d^{2-\epsilon})$ for any $\epsilon$. Therefore argument (ii) holds.
\\

\noindent\emph{Proof of argument (iii):}
We need to compare the compressed representation of the true deletion pattern $\mathbf{D}$ and that of the reconstructed deletion pattern $\widehat{\mathbf{D}}$ generated by the greedy alignment algorithm. We introduce a sequence $\mathbf{\Delta}=(\Delta_0, \Delta_1, \ldots, \Delta_n)$ to indicate the difference between $\mathbf{D}$ and $\widehat{\mathbf{D}}$.

Let $\Delta_0:=0$. For $i=1, 2, \ldots, n$, let $\Delta_i:=\Delta_{i-1}+D_i-\widehat{D}_{i}$. The condition $\Delta_i=0$ means that the greedy alignment algorithm is aligning $Y_i$ to the correct bit in $\mathbf{X}$. Given $\Delta_{i-1}$ and $D_i$, the value of $\Delta_i$ is as follows.
\begin{enumerate}
  \item If $\Delta_{i-1}=0$ and $D_i=0$, then $\widehat{D}_i$ must be $0$ and hence $\Delta_i=0$.
  \item If $\Delta_{i-1}=0$ and $D_i=1$, then $\widehat{D}_i = Y_i \oplus Y_j$, where $Y_j$ is the next undeleted bit. Since $\mathbf{Y} \sim$ iid Bernoulli$(p)$, $\widehat{D}_i \sim$ Bernoulli$(2p(1-p))$. Therefore $\Delta_i \sim$ Bernoulli$(1-2p+2p^2)$.
  \item If $\Delta_{i-1} \neq 0$, either $D_i=0$ or $D_i=1$, we have $\widehat{D}_i = Y_i \oplus Y_j$, where $Y_j$ is the next undeleted bit. Therefore $\widehat{D}_i \sim$ Bernoulli$(2p(1-p))$. Therefore $\Delta_i = \Delta_{i-1} + D_i - \widehat{D}_i$ where $D_i$ and $\widehat{D}_i$ are independent, $D_i\sim$ Bernoulli$(d)$ and $\widehat{D}_i \sim$ Bernoulli$(2p(1-p))$.
\end{enumerate}

Therefore $\mathbf{\Delta}$ is a first order Markov chain with the following transition probabilities:
$\pP(\Delta_i=1|\Delta_{i-1}=0)=1-\pP(\Delta_i=0|\Delta_{i-1}=0)=d(1-2p+2p^2)$. For $k \neq 0$,
$\pP(\Delta_i=k+1|\Delta_{i-1}=k)=d(1-2p+2p^2)$, $\pP(\Delta_i=k|\Delta_{i-1}=k)=2p(1-p)$, and $\pP(\Delta_i=k-1|\Delta_{i-1}=k)=(1-d)(1-2p+2p^2)$.
An important property of $\mathbf{\Delta}$ is that, when $d \ll 1$, starting from an arbitrary state, the Markov chain returns to the state $0$ in $O(1)$ steps on average. Therefore if the output of the greedy alignment algorithm disagrees with the true deletion pattern at some symbol, they will come back to an agreement in $O(1)$ steps.

When we read $\mathbf{D}$ and $\mathbf{Y}$ from left to right, if there is a deletion ($D_i=1$ for some $i$) and the run in $\mathbf{Y}$ that follows the run containing the deletion is not completely deleted, then the greedy alignment algorithm can locate the deletion in the correct run. For example, if $\mathbf{Y}=(0,0,1,0)$ and $\mathbf{D}=(1,0,0,0)$, the first `$0$' is deleted. The algorithm will generate $\widehat{\mathbf{D}}=(0,1,0,0)$, locating the deletion at the second bit, which is in the same run as the first bit. Since the compressed representations $\mathbf{V}$ and $\widehat{\mathbf{V}}$ are only about the locations of deletions up to runs, the corresponding entries in $\mathbf{V}$ and $\widehat{\mathbf{V}}$ related to this deletion are identical.

If there is a deletion, and the run in $\mathbf{Y}$ that follows the run containing the deletion is completely deleted, then the greedy alignment will locate some deletions in wrong runs. For example,  if $\mathbf{Y}=(0,0,1,0)$ and $\mathbf{D}=(1,0,1,0)$, the first `$0$' and the `$1$' are deleted. The algorithm will generate $\widehat{\mathbf{D}}=(0,0,1,1)$, locating the deletion of `$0$' incorrectly in the third run instead of the first run. Since such an event requires at least two deletions in consecutive runs, it occurs with probability $O(d^2)$. Since ${\mathbf{D}}$ and $\widehat{\mathbf{D}}$ will return to an agreement in $O(1)$ steps, with high probability, $n \cdot O(d^2)$ deletions may be placed in wrong runs by the greedy alignment algorithm throughout the sequence. Therefore up to $n \cdot O(d^2)$ entries of $\mathbf{V}$ and $\widehat{\mathbf{V}}$ can be different. Hence the entropy of the component-wise difference is $H(\mathbf{V}-\widehat{\mathbf{V}})=n \cdot O(d^2)$.

Therefore $|H(\widehat{\mathbf{V}})-H(\mathbf{V})|/n = |H(\widehat{\mathbf{V}}|\mathbf{V})-H(\mathbf{V}|\widehat{\mathbf{V}})|/n \leq 2 H(\widehat{\mathbf{V}}-\mathbf{V})/n=O(d^2)$, which completes the proof of argument (iii) and Theorem~\ref{thm:gapiid}.


\section{Proof of Theorem~\ref{thm:gapgeneral}}\label{app:proofthmgeneral}

In this appendix, let $\widehat{\mathbf{V}}$ denote $(\widehat{\mathbf {V}}^{ins},\widehat{\mathbf {V}}^{burst}, \widehat{\mathbf {V}}^{sub}, \widehat{\mathbf {V}}^{del})$.
Let $\mathbf{V}=({\mathbf {V}}^{ins},{\mathbf {V}}^{burst}, {\mathbf {V}}^{sub}, {\mathbf {V}}^{del})$ denote the corresponding description of the isolated insertions, bursty insertions, substitutions and deletions if the underlying sources
$\mathbf{Z}_X$, $\mathbf{Z}_Y$, $\mathbf{D}_X$ and $\mathbf{D}_Y$ are used. Note that the entries where both $\mathbf{D}_X$ and $\mathbf{D}_Y$ specify deletions are not considered as edits at all. The probability that such an entry occurs is $O(d^2)$.

As in the proof of Theorem~\ref{thm:gapiid}, we have
\begin{eqnarray*}
H(\mathbf X|\mathbf Y) &=& H(\mathbf X, \mathbf V|\mathbf Y) -
H(\mathbf V|\mathbf X, \mathbf Y)\\
&{=}& H(\mathbf V|\mathbf Y) -
H(\mathbf V|\mathbf X, \mathbf Y)\\
&=& H(\mathbf V) - I(\mathbf V;\mathbf Y)- H(\mathbf V|\mathbf X,
\mathbf Y),\\
&=& H(\widehat{\mathbf{V}}) - I(\mathbf V;\mathbf Y)- H(\mathbf V|\mathbf X,
\mathbf Y)-(H(\widehat{\mathbf{V}})-H(\mathbf{V})),
\end{eqnarray*}
We will prove the following three arguments: (i)$\lim_{n\rightarrow \infty} I(\mathbf {V};\mathbf {Y})/n=  O(d^{2-\epsilon})$ for any $\epsilon >0$,
(ii) $\lim_{n\rightarrow \infty} I(\mathbf V|\mathbf X, \mathbf Y)/n = O(d^{2-\epsilon})$ for any $\epsilon >0$, and (iii)
$\lim_{n\rightarrow \infty} |H(\widehat{\mathbf{V}})-H(\mathbf{V})|/n = O(d^{2-\epsilon})$ for any $\epsilon >0$.
\\

\noindent\emph{Proof of argument (i):}
$I(\mathbf {V};\mathbf {Y}) = I(\mathbf {V}^{ins};\mathbf {Y})+ I(\mathbf {V}^{burst};\mathbf {Y}|\mathbf {V}^{ins})+ I(\mathbf {V}^{sub};\mathbf {Y}|\mathbf {V}^{ins},\mathbf {V}^{burst})+I(\mathbf {V}^{del};\mathbf {Y}|\mathbf {V}^{ins},\mathbf {V}^{burst},\mathbf {V}^{sub})$. Due to the same reason as in the proof of argument (i) in Appendix~\ref{app:proofthm}, $I(\mathbf {V}^{ins};\mathbf {Y})=o(n)$. Since the bursty insertion appears with probability $O(d^2)$, $I(\mathbf {V}^{burst};\mathbf {Y}|\mathbf {V}^{ins})\leq H(\mathbf {V}^{burst}) = n\cdot O(d^{2-\epsilon})$. Since the substitutions represented by $V^{sub}$ are independent of $(\mathbf{Y},\mathbf{D}_Y,\mathbf{D}_X)$,  $I(\mathbf {V}^{sub};\mathbf {Y}|\mathbf {V}^{ins},\mathbf {V}^{burst})=0$. Due to the same reason as in the proof of argument (i) in Appendix~\ref{app:proofthm}, $I(\mathbf {V}^{del};\mathbf {Y}|\mathbf {V}^{ins},\mathbf {V}^{burst},\mathbf {V}^{sub})=0(n)$. Combining these four terms we have proved argument (i).
\\


\noindent\emph{Proof of argument (ii):}
The sequences $\mathbf{Z}_X$, $\mathbf{Z}_Y$, $\mathbf{D}_X$ and $\mathbf{D}_Y$ imply edits including insertions, deletions and substitutions.
Let us define the neighborhood of an edit as follows. The neighborhood of a substitution at position $i$ is the substitution together with the first run starting at position $(i+1)$, and the first bit of the second run. For example, when $\mathbf{Z}_X=(0,1,1,0,0)$, $\mathbf{Z}_Y=(1,1,1,0,0)$, there is a substitution at position $i=1$, the neighborhood of which consists of the first four bits. The neighborhood of a deletion in $\mathbf{D}_X$ at position $i$ is the run in $\mathbf{Z}_Y$ that contains the deletion, which ends at position $j$, together with positions $j+1, \ldots, j+k$, where $k$ is the smallest integer satisfying $k\geq 2$ and $Z_{Y,j+k}\neq Z_{Y,j+k-2}$. For example, $\mathbf{Z}_X=\mathbf{Z}_Y=(1,1,0,1,0,0)$, $\mathbf{D}_X=(1,0,0,0,0,0)$, there is a deletion at position $i=1$. The run containing the deletion ends at position $j=2$, and $k=4$. Therefore the neighborhood of this deletion consists of all six bits. The neighborhood of a deletion in $\mathbf{D}_Y$ is similarly defined. The concept of neighborhood is plays the same role as the ``extended run'' in the proof of argument (ii) in Appendix~\ref{app:proofthm}, because knowing that there is no other edit within the neighborhood of the first edit, without any ambiguity, the first edit can be located in the correct run if the edit is a deletion or insertion, and can be located precisely if it is a substitution.

When an edit appears and another edit appears within the neighborhood of the first edit, we call this neighborhood atypical. Let $\mathbf{Z}_X^*$, $\mathbf{Z}_Y^*$, $\mathbf{D}_X^*$ and $\mathbf{D}_Y^*$ be the sequences that are identical to $\mathbf{Z}_X$, $\mathbf{Z}_Y$, $\mathbf{D}_X$ and $\mathbf{D}_Y$ in the atypical neighborhoods, and take the value `$*$' otherwise. Thus the sequences are divided by the atypical neighborhoods into
$K$ segments of  `$*$'s. Let $C_{i}^{ins}$, $C_{i}^{sub}$, and $C_{i}^{del}$ be the numbers of insertions, substitutions and deletions in the $i$-th run, respectively. Let $\mathbf{C}:= (C_1^{ins}, C_1^{sub}, C_1^{del},\ldots, C_K^{ins}, C_K^{sub}, C_K^{del})$.

With the help of $\mathbf{Z}_X^*$, $\mathbf{Z}_Y^*$, $\mathbf{D}_X^*$ and $\mathbf{D}_Y^*$ and $\mathbf{C}$, aligning $\mathbf{X}$ and $\mathbf{Y}$ becomes easier. The atypical neighborhoods divide the whole sequences into $K$ segments. One can locate $K$ segments in $\mathbf{X}$ and $\mathbf{Y}$, each of which corresponds to a run of `$*$'s in $\mathbf{Z}_X^*$, $\mathbf{Z}_Y^*$, $\mathbf{D}_X^*$ and $\mathbf{D}_Y^*$. Within each segment, there are no longer atypical neighborhoods, and the edits can be located in the correct runs for insertions and deletions and can be located precisely for substitutions. Therefore $H(\mathbf{V}|\mathbf{X}, \mathbf{Y}, \mathbf{Z}_X^*, \mathbf{Z}_Y^*, \mathbf{D}_X^*, \mathbf{D}_Y^*, \mathbf{C})=0$, which implies that $H(\mathbf{V}|\mathbf{X}, \mathbf{Y})\leq H(\mathbf{Z}_X^*, \mathbf{Z}_Y^*, \mathbf{D}_X^*, \mathbf{D}_Y^*, \mathbf{C})$. Since an atypical neighborhood appears with probability $O(d^2)$, $\eE[K]/n = O(d^2)$ and $H(\mathbf{Z}_X^*, \mathbf{Z}_Y^*, \mathbf{D}_X^*, \mathbf{D}_Y^*, \mathbf{C})/n = O(d^{2-\epsilon})$ for any $\epsilon$. Therefore argument (ii) holds.
\\

\noindent\emph{Proof of argument (iii):}
Stage 1) of the algorithm specified in Section~\ref{subsec:algogeneralcase} generates a reconstructed alignment with the minimum number of edits, which can be compared with the original alignment specified by $\mathbf{Z}_X$, $\mathbf{Z}_Y$, $\mathbf{D}_X$ and $\mathbf{D}_Y$.

For each edit in the original editing process, if there is no other edit in its neighborhood, the reconstructed alignment must locate the correct type of edit within the correct run for if the edit is an insertion or a deletion, or at the correct position if the edit is a substitution. Otherwise the erroneous alignment leads to at least another edit in the neighborhood, which violates the assumption that the reconstructed alignment has the minimum number of edits.

If there is at least another edit in the neighborhood of the previous edit, so that the neighborhood is atypical, the reconstructed alignment is not guaranteed to be the same as the original alignment. Since such event occurs with the probability in the order of $O(d^2)$, the number of atypical neighborhoods is in the order of $n\cdot O(d^2)$. Therefore $\widehat{\mathbf{V}}$ and ${\mathbf{V}}$ differ by no more than $n\cdot O(d^2)$ entries. Therefore argument (iii) holds.


\footnotesize

\bibliography{newbibfile}

\begin{thebibliography}{10}
\providecommand{\url}[1]{#1}
\csname url@samestyle\endcsname
\providecommand{\newblock}{\relax}
\providecommand{\bibinfo}[2]{#2}
\providecommand{\BIBentrySTDinterwordspacing}{\spaceskip=0pt\relax}
\providecommand{\BIBentryALTinterwordstretchfactor}{4}
\providecommand{\BIBentryALTinterwordspacing}{\spaceskip=\fontdimen2\font plus
\BIBentryALTinterwordstretchfactor\fontdimen3\font minus
  \fontdimen4\font\relax}
\providecommand{\BIBforeignlanguage}[2]{{%
\expandafter\ifx\csname l@#1\endcsname\relax
\typeout{** WARNING: IEEEtran.bst: No hyphenation pattern has been}%
\typeout{** loaded for the language `#1'. Using the pattern for}%
\typeout{** the default language instead.}%
\else
\language=\csname l@#1\endcsname
\fi
#2}}
\providecommand{\BIBdecl}{\relax}
\BIBdecl

\bibitem{ISIT11}
{N.~Ma and K.~Ramchandran and D.~Tse}, ``{Efficient file synchronization: A
  distributed source coding approach},'' in \emph{Proc.~IEEE
  Int.~Symp.~Information~Theory}, {St. Petersburg, Russia}, {{Jul.~31--Aug.~5}}
  2011, pp. 583--587.

\bibitem{Leven}
V.~L. Levenshtein, ``Binary codes capable of correcting deletions, insertions
  and reversals,'' \emph{Doklady Akademii Nauk SSSR}, vol. 163, no.~4, pp.
  845--848, 1965.

\bibitem{OrlitskyDeletion}
A.~Orlitsky and K.~Viswanathan, ``One-way communication and error-correcting
  codes,'' \emph{IEEE Trans.~Inf.~Theory}, vol.~49, no.~7, pp. 1781--1788,
  2003.

\bibitem{Trachtenberg06}
S.~Agarwal, V.~Chauhan, and A.~Trachtenberg, ``Bandwidth efficient string
  reconciliation using puzzles,'' \emph{Parallel and Distributed Systems, IEEE
  Transactions on}, vol.~17, no.~11, pp. 1217 --1225, nov. 2006.

\bibitem{venkataramanan-interactive}
R.~Venkataramanan, H.~Zhang, and K.~Ramchandran, ``{Interactive Low-complexity
  Codes for Synchronization from Deletions and Insertions}.''

\bibitem{TridgellRsync}
A.~Tridgell and P.~Mackerras, ``The rsync algorithm,'' aNU Technical report,
  TR-CS-96-05, Jun 1996.

\bibitem{vsync}
H.~Zhang, C.~Yeo, and K.~Ramchandran, ``$\mbox{VSYNC}$: a novel video file
  synchronization protocol,'' \emph{ACM Multimedia}, pp. 757--760, 2008.

\bibitem{Verdu_universal}
H.~Cai, S.~Kulkarni, and S.~Verdu, ``An algorithm for universal lossless
  compression with side information,'' \emph{Information Theory, IEEE
  Transactions on}, vol.~52, no.~9, pp. 4008 --4016, sept. 2006.

\bibitem{Brandon}
M.~C. Brandon, D.~C. Wallace, and P.~Baldi, ``Data structures and compression
  algorithms for genomic sequence data,'' \emph{Bioinformatics}, vol.~25,
  no.~14, pp. 1731--1738, 2009.

\bibitem{MitzSurvey}
M.~Mitzenmacher, ``A survey of results for deletion channels and related
  synchronization channels,'' \emph{Probability Surveys}, vol.~6, pp. 1--33,
  2009.

\bibitem{maxrun}
M.~F. Schilling, ``The longest run of heads,'' \emph{The College Mathematics
  Journal}, vol.~21, no.~3, pp. 196--207, 1990.

\bibitem{nwalign}
S.~B. Needleman and C.~D. Wunsch, ``A general method applicable to the search
  for similarities in the amino acid sequence of two proteins,'' \emph{Journal
  of Molecular Biology}, vol.~48, no.~3, pp. 443--453, 1970.

\bibitem{MontanariISIT10}
{Y.~Kanoria and A.~Montanari}, ``{On the deletion channel with small deletion
  probability},'' in \emph{Proc.~IEEE Int.~Symp.~Information~Theory}, {Austin,
  Texas}, {{Jul.~13--18,}} 2010, pp. 1002--1006.

\end{thebibliography}
\end{document}